\documentclass{sigchi}

\usepackage{balance}       
\usepackage{graphics}      
\usepackage[T1]{fontenc}   
\usepackage{txfonts}
\usepackage{mathptmx}
\usepackage[pdflang={en-US},pdftex,]{hyperref}
\usepackage{color}
\usepackage{booktabs}
\usepackage{textcomp}
\usepackage{enumitem}

\usepackage{microtype}        
\usepackage{ccicons}          

\usepackage{todonotes}

\def\plaintitle{Comment Ranking Diversification in Forum Discussions}
\def\plainauthor{Curtis G. Northcutt, Kimberly A. Leon, Naichun Chen}
\def\plainkeywords{ranking diversification; discussion forum; online courses; information retrieval; search; embeddings}

\makeatletter
\def\url@leostyle{%
  \@ifundefined{selectfont}{
    \def\UrlFont{\sf}
  }{
    \def\UrlFont{\small\bf\ttfamily}
  }}
\makeatother
\def\pprw{8.5in}
\def\pprh{11in}

\definecolor{linkColor}{RGB}{6,125,233}
\hypersetup{%
  pdftitle={\plaintitle},
 pdfauthor={\plainauthor},
  pdfkeywords={\plainkeywords},
  pdfdisplaydoctitle=true, 
  bookmarksnumbered,
  pdfstartview={FitH},
  colorlinks,
  citecolor=black,
  filecolor=black,
  linkcolor=black,
  urlcolor=linkColor,
  breaklinks=true,
  hypertexnames=false
}

\begin{document}

\CopyrightYear{2017} 
\setcopyright{acmlicensed}
\conferenceinfo{Published in L@S 2017,}{April 20 - 21, 2017, Cambridge, MA, USA}
\doi{http://dx.doi.org/10.1145/3051457.3054016}

\title{\plaintitle}

\numberofauthors{1}
\author{%
  \alignauthor{Curtis G. Northcutt, Kimberly A. Leon, Naichun Chen\\
    \affaddr{Massachusetts Institute of Technology}\\
    \affaddr{Cambridge, MA, USA}\\
    \email{\{cgn, kimleon, naichun\}@mit.edu}}\\
}

\maketitle

\begin{abstract}
    Viewing consumption of discussion forums with hundreds or more comments depends on ranking because most users only view top-ranked comments. When comments are ranked by an ordered score (e.g. number of replies or up-votes) without adjusting for semantic similarity of near-ranked comments, top-ranked comments are more likely to emphasize the majority opinion and incur redundancy. In this paper, we propose a top $K$ comment diversification re-ranking model using Maximal Marginal Relevance (MMR) and evaluate its impact in three categories: (1) semantic diversity, (2) inclusion of the semantics of lower-ranked comments, and (3) redundancy, within the context of a HarvardX course discussion forum. We conducted a double-blind, small-scale evaluation experiment requiring subjects to select between the top 5 comments of a diversified ranking and a baseline ranking ordered by score. For three subjects, across 100 trials, subjects selected the diversified (75\% score, 25\% diversification) ranking as significantly (1) more diverse, (2) more inclusive, and (3) less redundant. Within each category, inter-rater reliability showed moderate consistency, with typical Cohen-Kappa scores near 0.2. Our findings suggest that our model improves (1) diversification, (2) inclusion, and (3) redundancy, among top $K$ ranked comments in online discussion forums. Code is open-sourced at \url{https://github.com/cgnorthcutt/forum-diversification}.
\end{abstract}



\section{Introduction}
Text ranking systems (e.g. Facebook post comments, Amazon product reviews, Reddit forums) are ubiquitous, yet many face a common problem. When posts (e.g. reviews or comments) are ranked primarily by text content and rating (e.g. like/unlike, $\uparrow$/$\downarrow$, +/-, number of replies, etc.), similar posts tend to receive similar scores. Moreover, higher ranking posts tend to exclusively represent the majority opinion, since there are more users in the majority group to up-vote posts sharing their sentiment. For large forums with thousands of posts, viewers may only be exposed to the majority opinion when they only view top-ranked posts. If the ground truth semantics of each comment were known a priori, comment scores could be normalized by the number of comments with similar semantics, avoiding this problem. Unfortuantely this is not the case. Instead, there are a multitude of techniques to approximate semantic similarity  \cite{mikolov2013distributed, dumais1988using, mueller2016siamese}.

\begin{figure}[ht]
    \begin{center}
    \vskip -.1in
    \centerline{\includegraphics[width=.95\columnwidth]{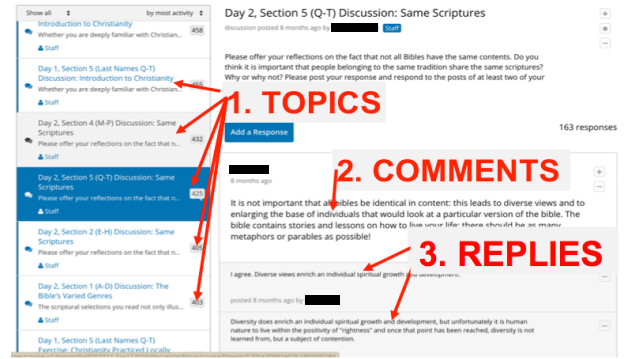}}
    \vskip -.2in
    \caption{An example topic used to illustrate the organization of an edX discussion forum. edX forums are organized hierarchically into topics $>$ comments $>$ replies. Our focus is the ranking of comments.}
    \label{fig:edx}
    \vskip -.1in
    \end{center}
\end{figure}

We consider the comment ranking diversity problem in the context of an online edX course, \emph{Harvardx Christianity Through Its Scriptures}, where increased visibility of the diversity of comments across thousands of learners may aid in debunking misconceptions held by the majority of forum respondents. edX forums are organized hierarchically into topics $>$ comments $>$ replies (an example topic is depicted in  Figure \ref{fig:edx}). Our focus is the ranking of comments and we use the number of replies as the score for each comment, although by default, edX comments are ranked chronologically.

In this paper, we develop an algorithm for forum comment ranking diversification using maximal marginal relevance (MMR) to linearly interpolate between the original \emph{relevance} ranking score of a comment and the \emph{diversity} of a comment with other high-ranked comments. We operationalize our notion of \emph{diversity} using a PCA + TFIDF model on all comments and evaluate our model using a blind experiment requiring subjects to compare our diversified ranking to a baseline relevance ranking.

\section{Background and Related Work}
The crux of diversification is a well-trained comment embedding model that accurately captures the semantic similarity between two documents. Text embedding is a well-studied problem at the word-level \cite{mikolov2013distributed} and document-level \cite{le2014distributed}. In this section, we consider increasingly complex methods for comment similarity, followed by methods for ranking documents and how it relates to diversification.

One of the simplest document embedding representations is TFIDF \cite{tfidf} which uses a "bag of words" (nBOW) counts model, normalized by word count per document frequency. Although TFIDF works well on some tasks \cite{aizawa2003information}, it ignores word ordering and suffers a performance loss for longer documents. TFIDF performs well when combined with matrix decomposition methods like PCA or LSA. More sophisticated approaches such as word2vec \cite{mikolov2013distributed}, LDA \cite{blei2003latent}, and Gated CNN \cite{barzilaysemi} offer classification accuracy improvements, but are task-specific. These models are compared in Table \ref{table:embedding_models}. A state-of-the-art (2016) LSTM similarity model uses a Siamese recurrent architecture to combine the word2vec embeddings of all words in a document, and trains using a Manhattan loss on the output of the two LSTMs \cite{mueller2016siamese}. Although this method would likely offer improvements, simpler models were sufficient for our task.

\begin{table}[ht]
\caption{A comparison of the comment embedding models evaluated in this study. Method symbols are abbreviated as: 
T=Topic, 
M=Matrix Factorization,
W=Local Window,
F=Frequency,
S=Semantic}
\label{table:embedding_models}
\centering
\resizebox{.9\columnwidth}{!}{%
\begin{tabular}{lcc}
\toprule
      \textbf{Model} & \textbf{Method} &  \textbf{Scaling Sensitivity}  
       \\
\midrule
      TFIDF &                F   &       False  \\    
     
      PCA + TFIDF &          M+S &       True  \\
      LSA + TFIDF &          M+S &       True  \\
      NMF + TFIDF &          M+S &       True  \\
      
      LDA + TFIDF &          T   &       False  \\
      
      Word2Vec + TFIDF &     W+S &       False \\
      Word2Vec + nBOW &      W+S &       False \\
      Gated CNN + TFIDF &    W+S &       False  \\

\bottomrule
\end{tabular}%
}
\end{table}

The task of forum comment ranking can be thought of as a search task, where common methods like PageRank \cite{page1999pagerank} and RankSVM \cite{duan2010empirical}) are used to identify the most relevant document for a given query. In our case, relevance is determined a priori by comment score, and instead our focus is diversification of this ranking. Diversification has been successfully applied to the task of online shopping \cite{chapelle2011intent}, with the task of reducing abandonment in shopping queries by providing a diversified selection of options. In this paper, we elect a more general approach, MMR \cite{carbonell1998use}, which we describe more in Section \ref{approach:mmr}.

\section{Technical Approach}
\label{technical-approach}

Our methodology consists of four ordered components: (1) Automated generation of gold data, (2) Evaluation of comment embedding models, (3) Implementing diversification in comment rankings, and (4) Measuring efficacy of diversification. We describe these components in the following sections.

\subsection{Dataset}
edX forums are organized hierarchically by topic $>$ comments $>$ replies as shown in Figure \ref{fig:edx}. We consider diversification at the comments level (within a single topic). In the context of this study, we focus on the comment rankings for topics in the forum discussions of an edX course, \emph{HarvardX: HDS3221.2x Christianity Through Its Scriptures}, obtained via web-scraping. Comment scores were set equal to the number of replies for each comment. Forum text was tokenized with stop-words removed and over 100,000 comments were analyzed.

\subsubsection{Automated Gold Data Generation}
We used a novel method to generate large gold datasets, without human labeling, by sampling comments across highly differing topics and generating a pairwise cosine similarity matrix for these comments. This matrix contains binary labels, a (1) if comments were taken from the same topic (Gold 1 pairs) or (0) if comments were taken from different topics (Gold 0 pairs). For exclusive sets of topics, we generated both train and test gold datasets to evaluate our selection of different comment embedding models discussed in  \ref{word-embedding-results}.

\subsection{Maximal-Marginal Relevance (MMR)} \label{approach:mmr}
MMR is an iterative algorithm, at each step selecting the comment which maximizes a modified score (Equation \ref{mmr}). 
\vskip -0.2in
\begin{equation} \label{mmr}
    \hat{s} := \lambda \cdot s - (1-\lambda) \cdot c
\end{equation}

A single parameter $\lambda$ adjusts the trade-off between the original comment score, $s$, and its maximum cosine similarity among all comments that have already been added to the new ranking, $c$, to produce the updated score, $s'$. For example, $\lambda = 1$ ranks entirely by score and $\lambda = 0$ selects maximally diverse comments irrespective of score. In this study, we evaluate two settings of the parameter, $\lambda = 0.75$ and $\lambda = 0.25$ in comparison with a baseline where $\lambda = 1$. 

\subsubsection{Comment Embedding Model Selection}
Diversification with MMR hinges on a comment embedding model that accurately captures the semantic similarity between two comments. Eight models were evaluated (Table \ref{table:embedding_models}).

Two evaluation metrics were used to compare these models. (1) The median quantile difference defined as the difference in average cosine similarity percentile rank (quantile) of Gold 1 pairs minus that of Gold 0 pairs. We recommend this metric as it is unbiased and captures relative ranking. (2) The accuracy of logistic regression using a given model's pairwise comment cosine similarity matrix as input and the gold binary labels as output. Our two metrics consistently ranked all models. 

Using the best performing model for these two metrics, comment similarity was computed using cosine similarity \cite{huang2008similarity}. In our case, the best model was PCA + TFIDF comment embeddings, as seen in Table \ref{embedding-results} in the Results section.

\subsection{MMR Evaluation Experiment}
\label{mmr-exp-description}

\begin{figure}[ht]
    \centering
    \includegraphics[width=.97\columnwidth]{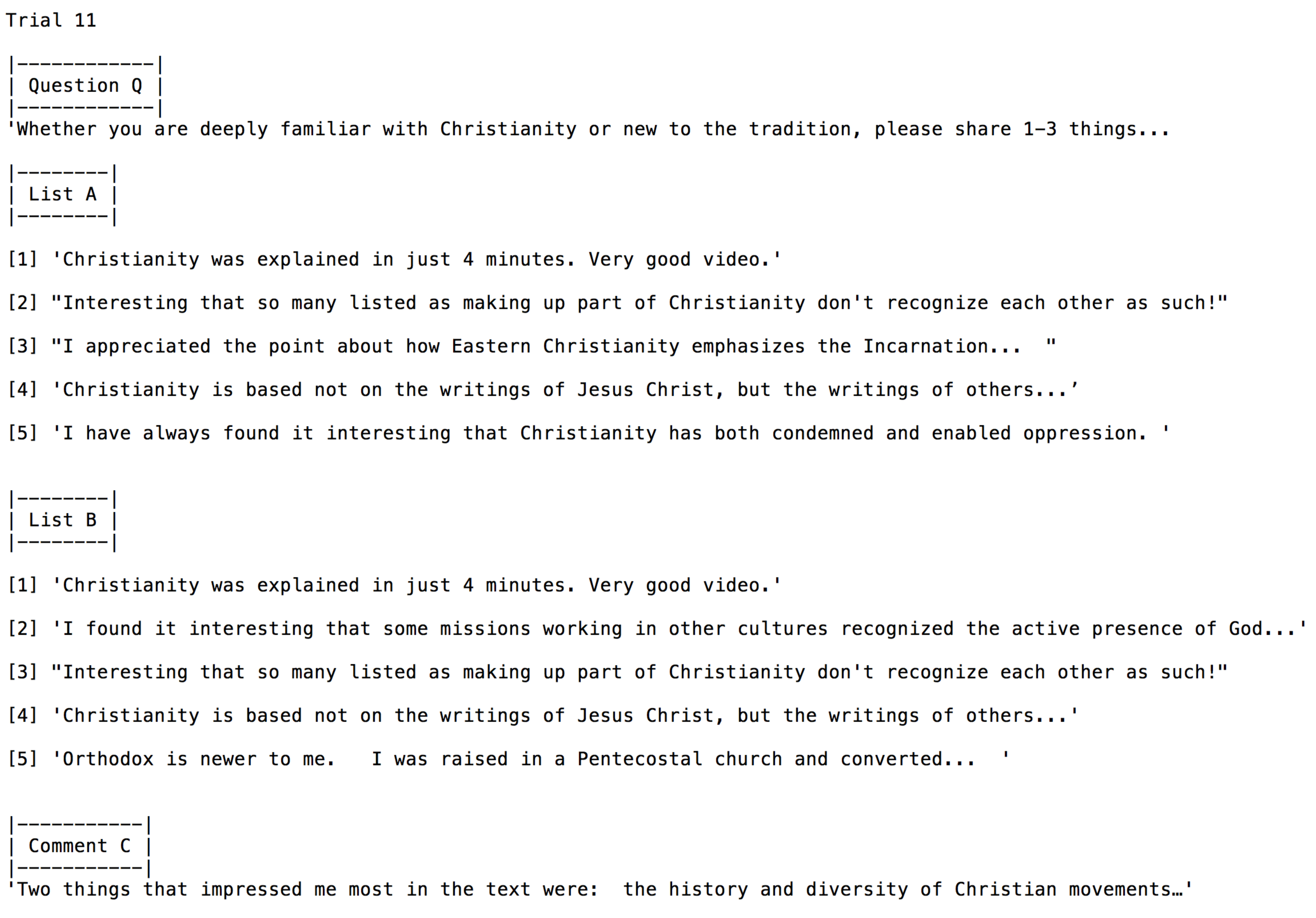}
    \caption{Example of a single trial in the MMR evaluation experiment. Each trial was presented to human subjects.}
    \label{fig:trial}
\end{figure}

Since the comment order for the course we experimented on is chronological, we used ordering by score (number of replies, $\lambda$ = 1) for our baseline ranking. We conducted a small-scale re-ranking evaluation experiment requiring subjects to choose among two unidentified ordered lists of comments: (1) the top 5 comments of our diversified ranking and (2) the top 5 comments of a baseline ranking ordered only by score, their true identities unknown. Three subjects evaluated 100 trials. The Cohen-Kappa score \cite{cohen1960} was used to measure inter-rater reliability. For each trial, subjects were presented with three tasks (an example trial is shown in Figure \ref{fig:trial}): 
\begin{enumerate} 
    \item The forum's topic question
    \item Two lists, A and B. One of these lists is the top five comments ordered by score (baseline). The other is the top five diversified (re-ranked) comments
    \item A random comment C from this forum not included in (2) where C's probability of being chosen was proportional to number of replies (higher rank = more likely to be chosen).
\end{enumerate}

Both the order in which lists A and B were shown to subjects and trial order were randomized to ensure the true labels for list A and B were unrecoverable within and across subjects. For each double-blind trial, each subject answered 3 questions: 

    1. \textbf{Inclusion Experiment}: Which list, A or B, has a comment that resembles the semantics of comment C?
    
    2. \textbf{Diversity Experiment}: Which list, A or B, best captures a diverse set of all potential answers to this question Q?
    
    3. \textbf{Redundancy Experiment}: Which list, A or B, contains more redundant comments?

If our comment embedding model accurately captures pairwise semantic similarity, we would expect the diversified ranking to be chosen more often for "inclusion" and "diversity", and less often for "redundancy".

Among the 100 trials for each subject, 75 trials used $\lambda= 0.25$ (ranked more by diversity) and 25 trials used $\lambda=0.75$ (ranked more by score). More trials were taken for $\lambda= 0.25$ to offset increased stochasticity when selecting low-scored (but diverse) comments. Neglecting comment score increases variation in ranking. Additional trials mitigated increased variance.

\section{Results and Discussion}
\label{results}

This section is divided into two parts. Since diversification relies on accurate semantic similarity scores, in Section \ref{word-embedding-results} we evaluate comment embedding models on our gold dataset. Then, in Section \ref{results:mmr}, we evaluate our model in a double-blind subject experiment comparing our diversified ranking against a baseline ranking ordered by score.

\begin{table}[ht]
\centering
\resizebox{.75\columnwidth}{!}{%
\begin{tabular}{lcc}
\toprule
                 \textbf{{\parbox{2cm}{Embedding Method}}} & \textbf{{\parbox{1.7cm}{Median Quantile Difference}}} &  \textbf{{\parbox{1.7cm}{Logistic Regression Accuracy}}} \\
\midrule
                     TFIDF &     0.338 &  0.841  \\    
         \
      \textbf{PCA + TFIDF} &  \textbf{0.434} & \textbf{0.867}\\
               LSA + TFIDF &     0.431 &  0.867  \\
               NMF + TFIDF &    0.416  &  0.861  \\                   
               LDA + TFIDF &    0.129  &  0.815  \\
               
          Word2Vec + TFIDF &     0.205 &  0.815  \\
          Word2Vec + nBOW  &     0.167 &  0.815  \\
         Gated CNN + TFIDF &     0.116 &  0.786  \\
\bottomrule
\end{tabular}%
}
\vskip .1in
\caption{Comparison of various comment embedding methods. Median quantile difference computes the difference in average cosine similarity rank (percentile) of Gold 1 pairs - Gold 0 pairs. Logistic regression predicts the accuracy of the gold labels trained using each model's pairwise cosine similarity matrix as input.} 
\label{embedding-results}
\end{table}

\subsection{Comment Embedding Models} \label{word-embedding-results}

For our task, word-level comment embedding methods (word2vec, Gated CNN, LDA) performed worse than a simple TFIDF vector representation alone, with a classical application of dimensionality reduction using PCA achieving highest accuracy on our gold dataset. Table \ref{embedding-results} captures the performance of different embedding models on our gold test set, for both median quantile difference and logistic regression accuracy. In the rest of this section, we discuss potential reasons for this. 

Comparing the use of the TFIDF embedding to the use of PCA and LSA affirms that there is benefit to employing dense embeddings. More unexpectedly, word2vec and Gated CNN, when combined with TFIDF, did not perform as well as TFIDF. A likely suspect is that our word2vec model was trained on the Google News corpus, which is a semantically different and much broader corpus than learner comments in an online course. As a result, word embeddings related to the course content were compressed into a smaller space relative to the broader embeddings of the model. 

Given that comments were on average 78 words in length, and "bag of words" ignores ordering and contextual information, it is less surprising that PCA and LSA outperformed nBOW and TFIDF models. As PCA offered a marginal performance improvement over LSA, PCA + TFIDF was chosen as our final comment embedding model.

\subsection{MMR Evaluation} \label{results:mmr}

\begin{figure}[ht]
    \begin{center}
    \centerline{\includegraphics[width=.8\columnwidth]{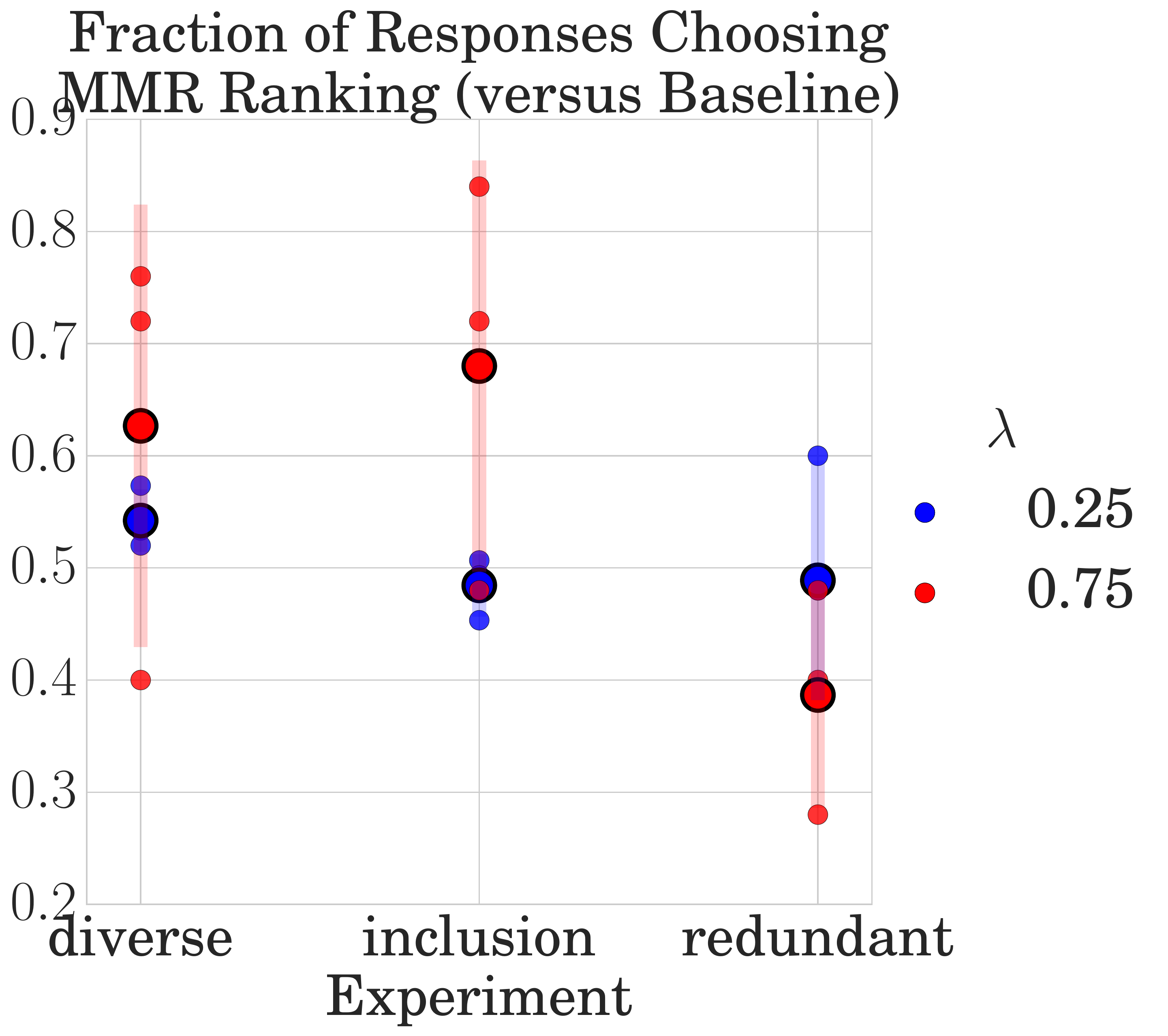}}
    \caption{Depicts the fraction of trials choosing the diversified (MMR) ranking for each $\lambda$, experiment pair. Higher values for the "diverse" and "inclusion" experiments and lower values for the "redundant" experiment suggest MMR's efficacy, with $\lambda = 0.75$ outperforming  $\lambda = 0.25$. The large, encircled points depict the means of each $\lambda$, experiment pair and the translucent bars depict the standard error of each mean. The smaller points depict individual rater scores. }
    \label{fig:mmr}
    \end{center}
\end{figure}

Table \ref{mmr-results} lists the results of the blind evaluation experiment. The fraction of subject responses selecting the diversified (MMR) ranking is depicted in Figure \ref{fig:mmr}. The MMR ranking with $\lambda = 0.75$ (ranked more by score) outperformed the baseline in every experiment (experiments are described in \ref{mmr-exp-description}), while rankings with $\lambda = 0.25$ (ranked more by diversity) did not perform significantly better or worse than the baseline.

\begin{table}[ht]
\centering
\resizebox{.9\columnwidth}{!}{
\begin{tabular}{llrrr}
\toprule
    $\mathbf{\lambda}$ & \textbf{Experiment} &  \textbf{Trials} & $\mathbf{\frac{Baseline}{Trials}}$ & $\mathbf{\frac{MMR}{Trials}}$ \\
\midrule
0.25 & inclusion &      225 &        0.52 &          0.48 \\
     & diverse &      225 &          0.46 &          0.54 \\
     & redundant &      225 &        0.51 &          0.49 \\
0.75 & inclusion &       75 &        0.32 &          0.68 \\
     & diverse &       75 &          0.37 &          0.63 \\
     & redundant &       75 &        0.61 &          0.39 \\
\bottomrule
\end{tabular} 
}
\caption{Depicts the aggregated subject counts of the blind evaluation experiment. For each ($\lambda$, experiment) group, the number of times either list was chosen is tallied. The two rightmost columns capture the normalized counts. The baseline ranking is generated with MMR and  $\lambda = 1$ (ranked only by score).} 
\label{mmr-results}
\end{table}

For moderate diversification ($\lambda = 0.75$), the MMR ranking was chosen significantly more often than the baseline ranking for both diversity and inclusion experiments, and significantly less often than the baseline for the redundancy experiment, suggesting our model mitigates redundancy and majority biases in the top $K$ comments. However, for extreme diversification ($\lambda = 0.25$) the fraction of responses choosing the MMR ranking was nearly $0.5$ (completely random when compared with the baseline ranking) across all three experiment groups. The cause is likely two fold. Firstly, ranking correlates with relevance, therefore, replacing more high-ranking comments with diverse, but lower-ranked (and less relevant) comments, may negatively impact all three experiments. Secondly, lower-ranked comments may be off-topic, lower quality, or harder to parse, leading to a simulated random choice.

\subsubsection{Reliability and Agreement Among Test Subjects}

Since only three subjects were included in our experiment, each evaluating 100 trials, we consider the inter-rater reliability among the three subjects to validate the consistency in our findings. Table \ref{mmr-results-irr} lists the Cohen's Kappa score for all pairs of subjects, for each experiment group. Although a small number of pairs were inconsistent, most showed moderate consistency.

\begin{table}[ht]
\centering
\resizebox{.75\columnwidth}{!}{
\begin{tabular}{llrr}
\toprule
        &        &    \textbf{other1} &    \textbf{other2} \\
\midrule
\textbf{diversity} & \textbf{subject 1} & -0.011 &  0.274 \\
        & \textbf{subject 2} & 0.179 &  0.274 \\
        & \textbf{subject 3} & -0.011 &  0.179 \\
\midrule
\textbf{inclusion} & \textbf{subject 1} & 0.034 &  0.147 \\
        & \textbf{subject 2} & 0.185 &  0.147 \\
        & \textbf{subject 3} & 0.034 &  0.185 \\
\midrule
\textbf{redundancy} & \textbf{subject 1} & -0.026 &  0.136 \\
        & \textbf{subject 2} & 0.211 &  0.136 \\
        & \textbf{subject 3} & -0.026 &  0.211 \\
\bottomrule
\end{tabular}
}
\caption{Cohen's Kappa pairwise inter-rater reliability scores.} 
\label{mmr-results-irr}
\end{table}

\section{Conclusion}
Discussion forums play a vital role in human interactions with online forums, yet due to large scale, comment rankings often suffer from majority biases and redundancy. The primary contributions of this paper are (1) design and evaluation of a top $K$ comment diversification re-ranking algorithm and (2) experimental evidence suggesting a significant increase in diversity and inclusion and decrease in redundancy when our algorithm is used to rank comments versus a baseline relevance ranking. We encourage large-scale commenting platforms, e.g. Facebook, edX, Reddit, etc., to consider the importance of ranking diversification on learning and user experience, and hope our findings inspire future consideration.

\section{Acknowledgements}
We graciously thank the reviewers who assisted in improving the manuscript, Y-Lan Boureau (Facebook AI Research) for her mentorship and suggestion of diversification using MMR with embeddings, and Regina Barzilay (MIT) for her guidance in model selection and framework.

\balance{}

\bibliographystyle{SIGCHI-Reference-Format}
\bibliography{paper}

\end{document}